\documentclass[]{emulateapj}

\usepackage{amsmath}
\usepackage{float}
\usepackage{booktabs}
\usepackage{hyperref}

\usepackage{color}

\begin{document}

\title{An Updated Galactic Double Neutron Star Merger Rate Based on Radio Pulsar Populations}

\author{Nihan Pol}
\author{Maura McLaughlin}
\author{Duncan R. Lorimer}
\affil{Department of Physics and Astronomy, West Virginia University, Morgantown, WV 26506-6315}
\affil{Center for Gravitational Waves and Cosmology, West Virginia University, Chestnut Ridge Research Building, Morgantown, West Virginia 26505}

\section{Updating the Galactic double neutron star merger rate}
    
    In this research note, we update the Galactic double neutron star (DNS) merger rate by including the new, highly eccentric DNS system J0509+3801 \citep{0509_disc_paper} which was recently discovered with the Green Bank North Celestial Cap survey \citep[GBNCC,][]{gbncc_survey}. We follow the same procedure described in \citet{my_dns_merger_rate}, but add the GBNCC survey to the list of surveys chosen for simulations (see Sec.~2.2 in \citet{my_dns_merger_rate}). Since it has not been measured, we assume a beaming correction factor of 4.6 for this pulsar, which is the average of the beaming correction factor of the known DNS systems \citep[see Sec.~2.5 of][]{my_dns_merger_rate}.
    
    Consequently, the new Galactic DNS merger rate is $\mathcal{R}_{\rm MW} = 37^{+24}_{-11}$~Myr$^{-1}$, where the errors represent 90\% confidence intervals. The small decrease (of $\sim 12\%$) in the total merger rate is due to the addition of the large GBNCC survey volume, but with the addition of only one merging DNS system to the total observed population. We can also see that the addition of a new DNS system results in tighter constraints on the Galactic DNS merger rate.
    
\section{Updating the merger detection rate for advanced LIGO}
    
    Similar to \citet{my_dns_merger_rate}, we can use this Galactic DNS merger rate to predict the number of DNS merger events that LIGO \citep{LIGO_detector_ref} will be able to detect. However, there was an error in the implementation of Eq.~15 in \citet{my_dns_merger_rate}. We used the range distance ($D_{\rm r}$) instead of the horizon distance ($D_{\rm h}$) in this equation, for a more conservative estimate, but have since realized that the derivation in \citet{extrapolate_to_get_LIGO_rate} already accounted for the reduction in LIGO’s sensitivity due to the orientation of the gravitational wave source with respect to the terrestrial detectors. Therefore, the horizon distance, which is a factor of 2.26 larger than the range distance \citep[$D_{\rm h} = 2.26 \times D_{\rm r}$,][]{use_range_not_horizon}, should be used with Eq.~15 in \citet{my_dns_merger_rate} in place of the range distance.
    
    Using the horizon distance in Eq.~15 of \citet{my_dns_merger_rate} along with the updated Galactic DNS merger rate results in a merger detection rate for LIGO,
    \begin{equation}
            \displaystyle \mathcal{R} = 1.9^{+1.2}_{-0.6} \times \left( \frac{D_{\rm r}}{100 \ \rm Mpc} \right)^3 \rm yr^{-1},
            \label{pml_rate}
    \end{equation}
    where $D_{\rm r}$ is the range distance. Using the LIGO O3 range distance of 130~Mpc \citep{LIGO_horizon_dist}, \textit{we predict that LIGO will detect anywhere between three and seven DNS mergers per year of observing at O3 sensitivity}.  
    
\begin{figure*}
        \centering
        \includegraphics[width = \textwidth]{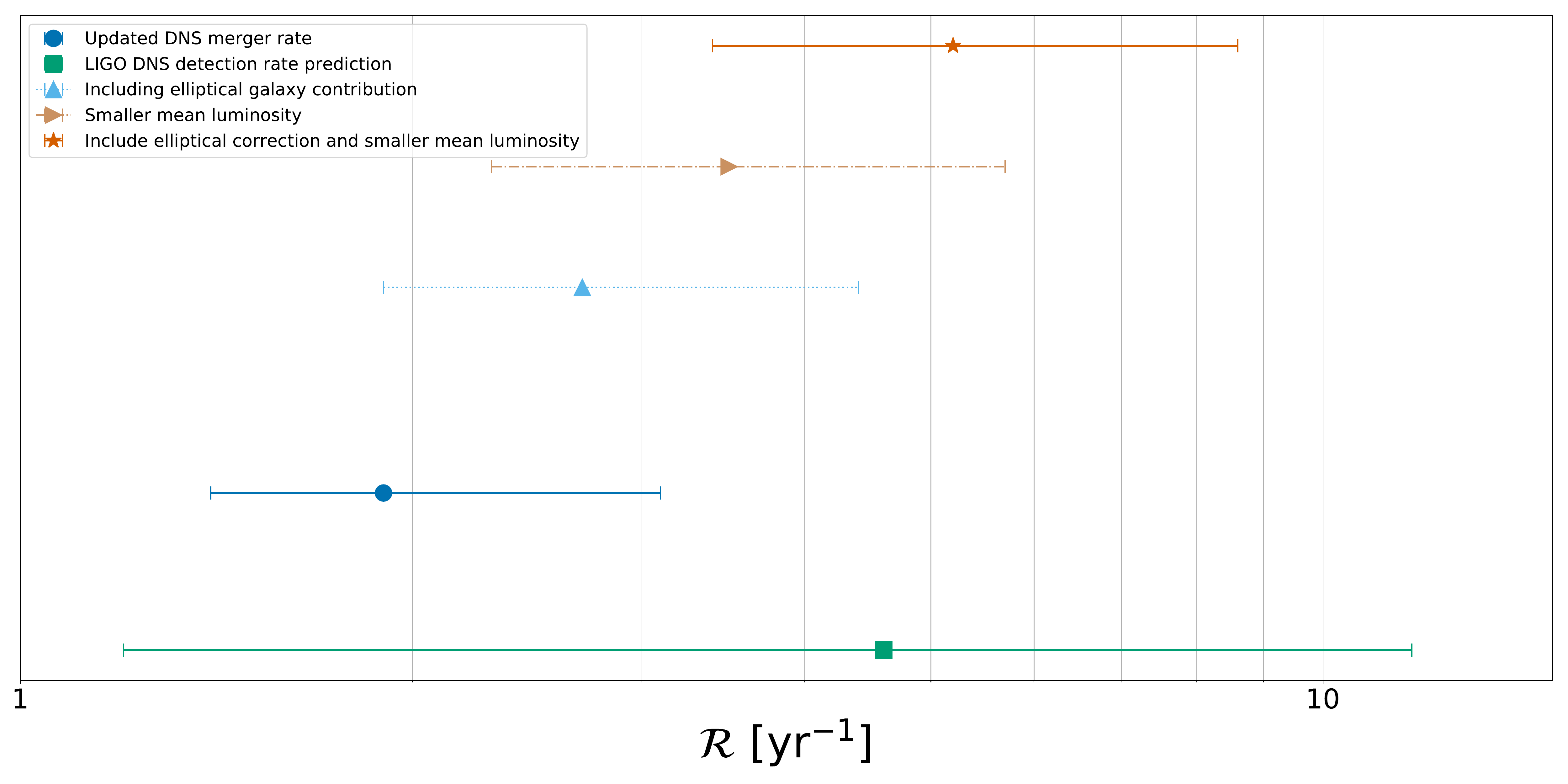}
        \caption{We compare the merger detection rate calculated using the observed Galactic DNS population (Eq.~\ref{pml_rate}) with the rate calculated by LIGO's second detection of a DNS merger (Eq.~\ref{LIGO_rate}). We also show the variation in the predicted merger detection rate due to an underlying pulsar population with a lower mean luminosity, as well as the effect of including the contribution of elliptical galaxies in the extrapolation of the Galactic merger rate to LIGO's observable volume \citep{extrapolate_to_get_LIGO_rate}. We also plot the modified merger detection rate that includes both the correction for elliptical galaxies and a fainter DNS population. The data required to make this figure is provided as a supplementary table in CSV format.}
        \label{compare_rates}
    \end{figure*}

    We can compare this merger detection rate derived from the observed Galactic DNS population to that calculated using LIGO's second observation of a DNS merger event, GW190425 \citep{second_dns_merger}. The DNS merger detection rate calculated using GW190425 and GW170817 \citep{second_dns_merger}, converted to the above units, is, 
    \begin{equation}
            \displaystyle \mathcal{R}_{\rm LIGO} = 4.6^{+7.1}_{-3.4} \times \left( \frac{D_{\rm r}}{100 \ \rm Mpc} \right)^3 \rm yr^{-1}.
            \label{LIGO_rate}
        \end{equation}
    We plot this merger rate together with the merger detection rate predicted using the Galactic DNS population in Fig.~\ref{compare_rates}. As in \citet{my_dns_merger_rate}, we also plot the merger detection rate predictions due to variations in the underlying pulsar luminosity distribution, as well as including the effect of inclusion of elliptical galaxies in the merger rate extrapolation \citep{use_range_not_horizon}.
    
    We conclude that the Galactic DNS merger detection rate is consistent with the merger detection rate calculated using gravitational wave detection of DNS mergers by LIGO. The number of alerts issued by LIGO in O3 for potential (i.e. unconfirmed) DNS mergers is also consistent with the predictions made using the Galactic DNS population.
    
\acknowledgements

We would like to thank Ilya Mandel for correcting our interpretation of Eq.~15 in \citet{my_dns_merger_rate} and for a number of very useful discussions that led to the results presented here.

NP and MAM are members of the NANOGrav Physics Frontiers Center (NSF PHY-1430284). MAM and DRL have additional support from NSF OIA-1458952 and DRL acknowledges support from the Research Corporation for Scientific Advancement and NSF AAG-1616042.

\bibliography{bibliography.bib}
\bibliographystyle{aasjournal}

\end{document}